\documentclass[pra,twocolumn,showpacs,preprintnumbers,
amsmath,amssymb,floatfix]{revtex4}
\usepackage{hyperref}
\usepackage[dvips]{epsfig}
\newcommand{\beq}{\begin{equation}}
\newcommand{\eeq}{\end{equation}} 
\newcommand{\beqa}{\begin{eqnarray}}
\newcommand{\eeqa}{\end{eqnarray}}
\newcommand{\ba}{\begin{array}}
\newcommand{\ea}{\end{array}}

\begin{document}
\title{Critical velocity, vortex shedding and drag in a 
unitary Fermi superfluid}
\author{F. Ancilotto$^{1,2}$, L. Salasnich$^{1}$, and F. Toigo$^{1,2}$} 
\affiliation{$^1$Dipartimento di Fisica e Astronomia 
"Galileo Galilei" and CNISM, Universit\`a di Padova, 
Via Marzolo 8, 35122 Padova, Italy \\
$^2$CNR-IOM Democritos, via Bonomea, 265 - 34136 Trieste, Italy} 

\begin{abstract} 
We study the real-time motion of a microscopic object  
in a cold Fermi gas at unitary conditions by 
using an extended Thomas-Fermi density functional approach.
We find that
spontaneous creation of singly quantized vortex-antivortex
pairs occurs as a critical velocity is exceeded,
which leads to a drag between the moving object and the Fermi gas. 
The resulting force is linear in the velocity for subsonic motion 
and becomes quadratic for supersonic motion.
\end{abstract} 


\maketitle

\section{\bf INTRODUCTION} 

A Fermi gas of atoms at unitary conditions
is characterized by a divergent 
s-wave scattering length \cite{bertsch}, 
which makes it a unique system, being at the same
time dilute and strongly interacting.
In this regime all scales 
associated with interactions disappear from the problem 
and the energy of the system is expected to be proportional to that 
of a non interacting fermions system.
The unitary Fermi gas (UFG) at low energies is known 
to be superfluid, and 
the most striking manifestation of its phase coherence
has been the experimental 
observation of vortices in a rotating UFG of $^6$Li atoms \cite{vort}.

Vortex dipoles in a Fermi gas were studied 
theoretically, within an extended Thomas-Fermi approach,
in the BCS limit \cite{gautam}.
Vortex solutions within the same 
approach have been studied in Ref. \cite{cetoli}, where
the rotation of the Fermi system was predicted
to generate a giant vortex in the presence of strong 
anharmonicity in the confining potential.
Recently \cite{science} 
a time-dependent Bogoliubov deGennes (BdG) approach
has been used to study the 3-D, real-time formation of
vortices in a UFG.
Surprisingly, one of the main conclusion in Ref. \cite{science}
is that the system remains superfluid even when stirred at
supercritical speed.
Single vortex solutions in the Ginzburg-Landau regime 
of a trapped superfluid Fermi gase have been studied
in Ref. \cite{rodriguez} and Ref. \cite{bruun}. 
In a dilute Fermionic superfluid 
a vortex state is characterized by a strong
density depletion along the vortex core. The depletion is
however not complete, according to the BdG
calculations of Refs. \cite{bulg-yu,sensarma}.

Recently  
it has been remarked \cite{bulgac1} that 
the superfluid unitary Fermi gas
can be efficiently described at zero temperature by 
phenomenological density functional
theory. Density functionals of different flavours 
have been proposed by different theoretical groups.
Bulgac and Yu have introduced a superfluid 
density functional based on a Bogoliubov-de Gennes 
approach to superfluid fermions \cite{bulgac2,bulgac3}. 
Papenbrock and Bhattacharyya \cite{papenbrock} 
have instead proposed a Kohn-Sham density 
functional with an effective mass 
to take into account nonlocality effects. 
Here we adopt instead the extended Thomas-Fermi functional of the 
UFG that we have proposed few years ago \cite{salasnich}
and which has been used recently to successfully 
address a number of properties of such system 
\cite{salasnich,best,Sala_Anci_Jos,anci3,salasnich-ratio}. 

Our extended Thomas-Fermi (ETF) density functional 
approach for the UFG \cite{salasnich,best} is based
on the following extended hydrodynamics equations:
\begin{eqnarray}
{\partial n \over \partial t} + {\boldsymbol 
\nabla} \cdot (n {\bf v}) = 0 \; , 
\label{hy1}
\\
m{\partial {\bf v} \over \partial t} + {\boldsymbol 
\nabla} [ {m\over 2} v^2 + U({\bf r,t})+
{ \xi} {\hbar^2 \over 2m} (3\pi^2)^{2/3} n^{2/3}\nonumber \\
-
\lambda {\hbar^2 \over 2 m} {(\nabla ^2 \sqrt{n})\over \sqrt{n}}] = {\bf 0} \; 
\label{hy2}
\end{eqnarray}
where $n({\bf r},t)$ is the time-dependent scalar density field 
and ${\bf v}({\bf r},t)$ the time-dependent velocity field.  
Here $U({\bf r,t})$ is the external potential and 
$\xi = 0.40$ and $\lambda = 1/4$ \cite{best,anci3}.
The above equations describe accurately various static 
and dynamical properties of the UFG. 
The term multiplied by $\lambda$ is found to be crucial 
to accurately describe surface effects, 
in particular in systems with a small number of atoms, 
where the Thomas-Fermi (local density) approximation 
fails \cite{salasnich}. 
We have shown that when fast dynamical processes occur and/or when 
shock waves come into play such term is necessary 
also in the large N limit \cite{anci3}, where 
results in quantitative agreement with 
experiments of real-time collision of strongly 
interacting Fermi gas clouds at unitarity \cite{thomas}
can be obtained \cite{anci3}.

We use this method here to study the motion 
of a microscopic 2-dimensional object 
with circular shape in a UFG and the 
associated process of vortex shedding, once the
object velocity exceeds a critical value.
Experimental realization of this geometry 
would imply, for instance,
moving a far-detuned laser beam 
through a trapped condensate.

\section{\bf CRITICAL VELOCITY AND DRAG IN THE UNITARY FERMI GAS} 

For a dilute Fermi gas the long wavelength elementary 
excitations are sound waves, and the Landau criterion
for the critical velocity $v_c=(\epsilon /p)_{min}$
for breakdown of superfluidity gives $v_c=c_1$.
However, for fluid flowing past an object
(or an atomic impurity as well), the local 
velocity near the object surface can become supersonic
even when the fluid velocity 
far from the object, $v_{\infty}$, is subsonic
(the maximum velocity of, e.g., a 2-dimensional flow
of an incompressible fluid past a circular obstacle is reached 
on the perimeter of the obstacle, where it  is $\sim 2v_{\infty}$).

An estimate for the critical velocity for a UFG
can be obtained as follows.
At stationary conditions 
Eq. (\ref {hy2}) provides the Bernoulli equation for the UFG:
(in the following 
$\alpha \equiv \xi {\hbar ^2 \over 2m}(3\pi ^2)^{2/3})$
\beq
-\lambda {\hbar^2 \over 2 m} {(\nabla ^2\sqrt{n})\over \sqrt{n}}
+\alpha n^{2/3}+U({\bf r})+{m\over 2} v^2=const
\label{bern}
\eeq

By assuming that the quantum term in the previous equation is 
negligible in the spirit of the long-wavelength limit approximation,
and calling $n_0$ the (uniform) density far from an impenetrable 
object, where $v\sim v_{\infty}$, one finds:
\beq
n({\bf r})=[n_0^{2/3}+{m\over 2\alpha}(v_\infty ^2 -v^2)]^{3/2}
\label{ncrit}
\eeq 
outside the object and $n=0$ within it (the interaction U between the 
object and the fluid only provides the excluded volume condition).
 
The above equation may be written in terms of the local 
\beq 
c_{loc}\equiv \sqrt{\xi \over 3}v_F({\bf r})=
\sqrt{2\alpha n({\bf r})^{2/3}/(3m)}
\label{cloc}
\eeq
and bulk
\beq
c_1\equiv \sqrt{\xi \over 3}v_{F,\infty}
=\sqrt{2\alpha n_0^{2/3}/(3m)}
\label{sound}
\eeq
speeds of sound as:
\beq
c_{loc}^2=c_1^2+(v_\infty ^2-v^2 )/3
\label{ncrit1}
\eeq
When $v\sim c_{loc}$ local instabilities develop,
leading to the release of vortices \cite{frisch}. 
The maximum local velocity of a fluid flowing past an 
impenetrable 
cylindrical object, normally to its axis, is
$v\sim 2v_\infty$ (on the surface of the cylinder
and tangent to it). 
Using this value in (\ref{ncrit1})
one can solve the equation for $v_\infty $, thus 
providing an approximate value for the
critical velocity of the fluid flowing past a
stationary object (or, equivalently,
of a moving object in the fluid at rest):
\beq
v_c= c_1/\sqrt{5}
\label{vcrit}
\eeq

This value is very similar to the one,
$v_c\sim \sqrt{2/11}c_1$ obtained 
for the BEC case, using a similar argument,
in Ref. \cite{frisch}.
For a spherical object the equatorial velocity 
is instead $v\sim 3v_\infty /2$, giving for the
critical velocity of a sphere moving in a UFG
$v_c= \sqrt {3/8}c_1$.

By combining the two equations (\ref{hy1}) and (\ref{hy2})
one can derive the equation for the 
momentum conservation (summation over repeated indices is 
implied):
\beq
\partial _t J_k + \partial _i T_{ik} +\rho \partial _k (U/m)=0
\label{momeq}
\eeq
where $J_k=\rho v_k$ is the supercurrent density
and the stress tensor $T_{ik}$ is defined as:
\begin{eqnarray}
T_{ik}\equiv \rho v_i v_k+\delta _{ik}({2\alpha \over 5m}\rho ^{5/3})
-{1\over 4} ({\hbar \over 2m})^2\rho \partial _i \partial _k ln(\rho)
\label{stress}
\end{eqnarray}
(to derive the above equation the following identity
has been used:
$\rho ^{-1}\partial _i(\rho \partial i \partial _k ln(\rho ))=
2\partial _k(\partial _i \partial _i \sqrt{\rho}/\sqrt{\rho})$).

Note that although in a superfluid there is no
frictional viscosity, nonetheless shear stress may arise from density
(pressure) gradients (third term in Eq. (\ref {stress}).
As a consequence, vortex formation and
drag are possible even with no viscosity.

The force 
exerted on a condensate by an obstacle moving through it
can be calculated from the rate of momentum transfer
to the fluid. By integration, one finds 
the drag force (per unit mass):
\beq
F_k=\partial _t \int _\Omega d\Omega J_k =
-\int _\Sigma d\Sigma n_i T_{ik}-\int _\Omega d\Omega \rho
\partial _k U
\label{drag}
\eeq

Here $\Sigma $ is the object external surface and 
${\bf n}$ is a unit vector directed along the outward normal.
In the case of an homogeneous flow past 
an impenetrable object only the first term contributes.
In the present case, however, where a partially penetrable
object is used (see the following) both contributions are present.

In our calculations we used  the full expression (\ref {stress}) for 
the stress tensor to compute the
drag (\ref {drag}). Notice that the quantum term in (\ref {stress}) is
expected to be negligible only when $v _\infty \gg (\hbar /mR)$,
where $R$ is the diameter of the moving object:
this is not the case here since these two terms are 
of comparable magnitude.

\section{\bf METHODS AND CALCULATIONS} 

By using a Madelung transformation, equations (\ref{hy1}) and (\ref{hy2}) 
can equivalently be written  
in the form of a time-dependent 
nonlinear Schr\"odinger equation (NLSE) \cite{salasnich}
involving the complex order parameter $\Psi({\bf r},t) = \sqrt{n({\bf r},t)}\
e^{i\theta({\bf r},t)}$:
\beq 
i \hbar {\partial \over \partial t} \Psi = \hat{H}\Psi
\eeq
where
\beq
\hat{H}\equiv -{\hbar^2 \over 4 m} \nabla^2 + 2 U({\bf r}-{\bf V}t) +
2 \alpha |\Psi|^{4/3} 
\label{nlse}
\eeq 

The link between the two descriptions is provided by 
the definition
\beq
{\bf v}({\bf r},t) \equiv {\hbar \over 2m}\nabla \theta({\bf r},t)
\eeq
for the velocity field associated with the phase
of the order parameter $\theta({\bf r},t)$. 
{\bf V} in Eq. (\ref {nlse}) is the (constant) velocity 
of the moving object. 

By means of a Galilean transformation to the reference frame moving with the
object (which will thus appear as stationary in our simulations), 
the NLSE to be solved becomes:
\beq
i \hbar {\partial \over \partial t} \Psi = \Big[\hat{H}+i\hbar {\bf V}
\cdot \nabla \Big]\Psi
\label{nlse1}
\eeq

For simplicity, we model the microscopic object inside the moving Fermi 
gas by means of a repulsive cylindrical "wall" and consider the flow 
motion perpendicular to it. Due to translational invariance along the 
axis of the cylinder (which we define as the $z$ axis), the problem 
thus reduces to finding the density and the velocity of the fluid in 
the $(x,y)$ plane.

We have numerically solved this 2-D NLSE equation to obtain the long-time 
dynamics of the fluid due to the motion of the object moving
along the $x$ direction.
From the time-dependent calculated density 
$n(x,y,t)$ and velocity ${\bf v}(x,y,t)$
we computed the drag acting on the object, 
according to (\ref {drag}).

Our simulation region is a square cell of side $a=3.2\,\mu m$.
We used a uniform mesh to represent the wavefunction $\Psi $, 
on $512\times 512 $ points in the $(x,y)$ plane.
We have used the Runge-Kutta-Gill fourth-order method \cite{rkg} 
to propagate in time the solutions of the NLSE. 
To accurately compute the spatial derivatives 
appearing in our NLSE, we used a 13-point finite-difference 
formula \cite{pi}. 
To avoid the outgoing sound waves and/or emitted vortices
to interfere with the fluid dynamics after being reflected on
on the grid boundaries, we use an exponential absorbing buffer
located in the periphery of the cell, as described in Ref. \cite{dafei}.
The waves can travel freely in the undamped region
(which occupy most of the simulation cell)
but are quickly attenuated as they enter the 
external region, thus preventing unwanted interferences
which might spoil the results.

In the following we take $m$  equal to
the mass of a $^6$Li atom.
We consider two different density values ("high density" and
"low density" in the following) for the UFG, namely 
$n_0=8600\,\mu m ^{-3}$ and 
$n_0=880\,\mu m ^{-3}$, corresponding to  
interparticle distances $d\sim 5\times 10^{-2}\, \mu m$
and $d\sim 0.1 \mu m$, respectively.

\begin{figure}
\epsfig{file=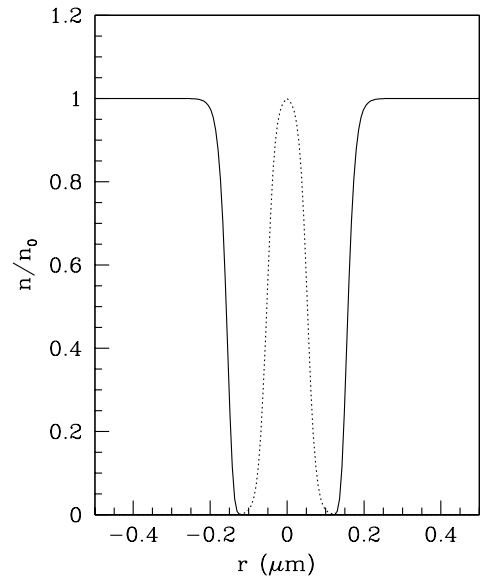,height=3.3 in,clip=}  
\caption{Equilibrium density profile 
at t=0. 
The dotted line represents (in arbitrary units) the repulsive potential 
due to the object.}
\label{fig1}
\end{figure}

At equilibrium, the repulsive cylindrical "wall"
 results in the 
formation of a circular cavity void of atoms.
The object potential $U$ and the 
associated initial density profile
are shown in Fig. (\ref {fig1}).

\section{\bf RESULTS AND DISCUSSION} 

We made a series of time-dependent calculations,
solving Eq. (\ref {nlse1}) for different values of the speed $V$,
and following the dynamics of the systems for several 
$\mu s$.
We find that there is a critical value $v_c$ (which will
be quantified in the following) for 
the object speed $V$ which 
separates two distinct regimes. When $V<v_c$ the 
fluid profile rapidly evolves with time into a
stationary configuration. Both the density 
and the velocity field for such final configuration
are fore-to-aft symmetric. This implies
that the drag force on the object is zero, which is
another version 
of the well-known D'Alembert paradox in classical fluids.

\begin{figure}
\epsfig{file=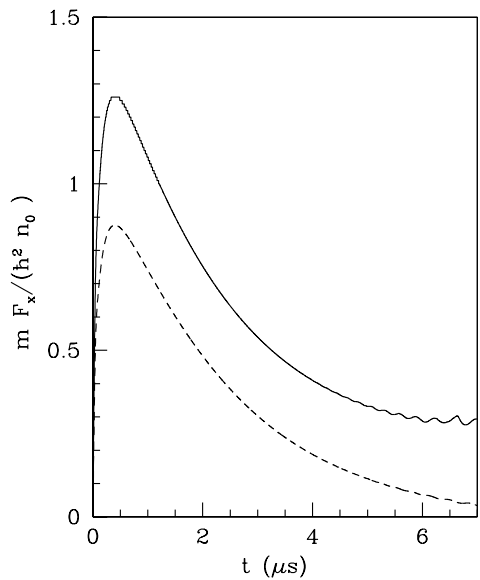,height=3.3 in,clip=}  
\caption{Drag force during the real-time evolution of
the system. Upper curve: $V>v_c$, lower curve: $V<v_c$.}
\label{fig7}
\end{figure}

In Fig. (\ref{fig7}) (lower curve) we show the calculated
drag force $F_x$ (computed using Eq. (\ref {drag})) 
as a function of time when $V<v_c$. After a transient where the 
fluid accomodates under the sudden acceleration of the
object, the drag goes eventually to zero when the symmetric
(stationary) configuration is reached. 

Above $v_c$, however, the Fermi gas dynamics 
changes dramatically: linear (antilinear)
vortices are spontaneously created in pairs close to the surface
of the object, together with the emission of sound waves.

\begin{figure}
\epsfig{file=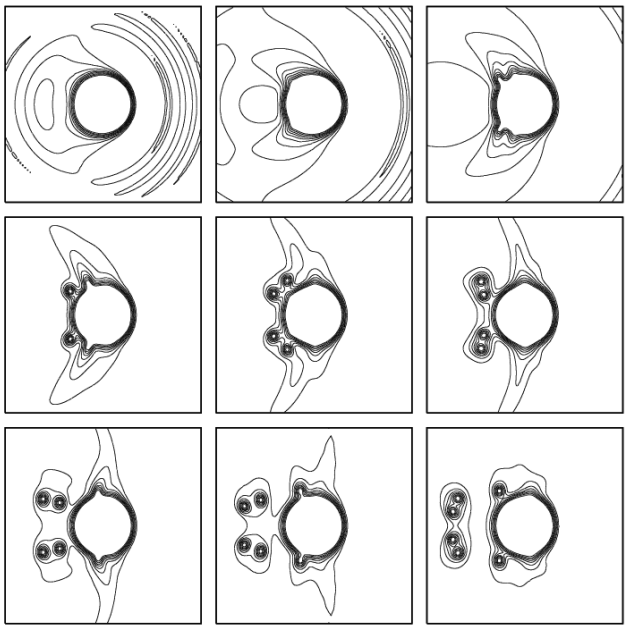,height=3.3 in,clip=}  
\caption{From left to right, top to bottom:
configurations at increasing times 
for $V/c_1 = 0.76$.
Only a portion of the simulation cell is displayed.}
\label{fig2}
\end{figure}

We show in Fig. (\ref{fig2}) a sequence of 
images taken during the real-time evolution of the system
when $V>v_c$.
The object is moving from left to right.
First a localized bow wave moving with supersonic velocity is emitted
in front of the cylinder and rapidly moves ahead.
This is the result of a "shock" wave produced by the 
sudden acceleration of the object (we recall that the initial 
state is with fluid at rest and the object moving
at constant velocity).

Then a vortex pair is emitted in the rear, trailing behind
(meaning that 
the pair velocity is lower than that of the
moving object).
The initial pair is soon overtaken by a pair emitted successively:
the vortex lines apparently form a temporarily 
bound state (but they will eventually split at later
times, not shown).

As shown in the upper curve of Fig. (\ref {fig7}), in this case  the 
instantaneous calculated
drag force
relaxes, after a transient, towards a finite nonzero value (with oscillations
that reflect the quasi-periodic emission of vortex pairs).

In the simulations of Fig. (\ref {fig2})
the velocity $V$ is greater than $v_c$, but still 
below the speed of sound $c_1$.

\begin{figure}
\epsfig{file=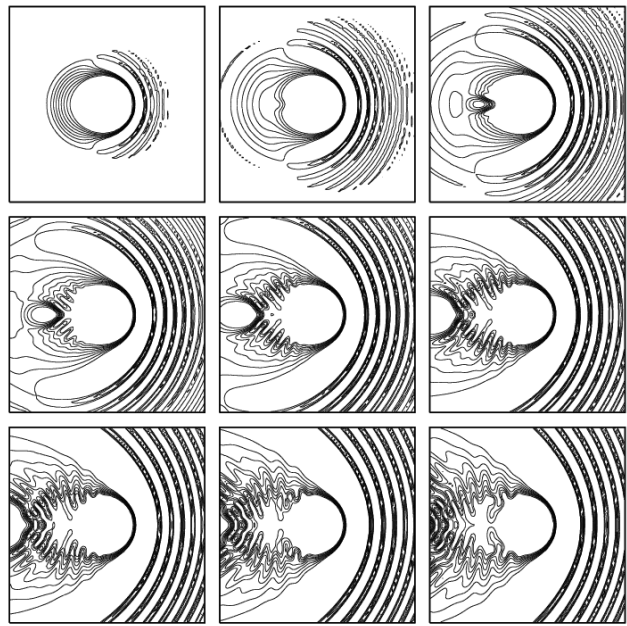,height=3.3 in,clip=}  
\caption{From left to right, top to bottom:
configurations at increasing times 
for velocity $V/c_1 = 1.44$.
Only a portion of the simulation cell is displayed.}
\label{fig3}
\end{figure}

The sequence shown in Fig. (\ref {fig3}) is instead obtained
for a supersonic motion of the object, $V/c_1 = 1.44$.
It appears that the vortex shedding 
frequency is considerably increased
with respect to the previous case (similarly to what happens for a BEC,
where the shedding frequency is \cite{winie1} $\propto V^2$.
Vortex-antivortex pairs are emitted in a semicontinuous way
on different part of the rear section of the object,
leading to parallel rows of connected vortical lines
that eventually decay into separated vortices.
Note also the appearance of a rather structured 
bow sound wave pattern moving along with the object.
As time proceeds, the wake region behind the 
object becomes turbulent due to the superposition 
of more and more vortices and sound waves.

In order to better characterize the vortex structure,
we have selected (from another simulation where 
$V$ is just above $v_c$, so that 
vortex pairs are emitted with low frequency and thus are
well separated from one another)
a configuration after a pair of vortex lines 
have been emitted and moved away from the 
object, and closely analyzed the vortex structure.
We find that the vortex is singly quantized,
and that the
two vortices of the pair have opposite polarization.
The velocity field (not shown) follows very closely
the ideal vortex velocity profile, 
$v(r)=\hbar /(2mr)$ (here $r=\sqrt{x^2+y^2}$).

\begin{figure}
\epsfig{file=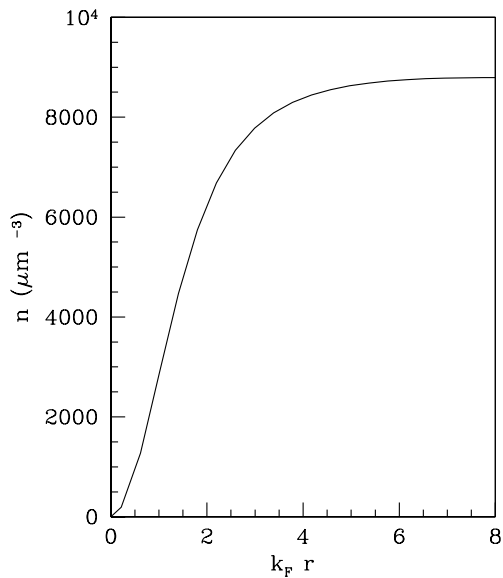,height=3.3 in,clip=}  
\caption{Density profile in the vortex core region.}
\label{fig5}
\end{figure}

The vortex line structure has an empty core,
as shown in Fig. (\ref {fig5}), while the core size scale
is set by $k_F^{-1}$.
This is in contrast with calculations based on Bogoliubov-De Gennes 
calculations \cite{sensarma,bulg-yu} 
where a partially filled core (between 0.2 and 0.3 $n_0$)
is predicted due to the presence of some normal liquid
coming from pair breaking in the core region, where the velocity is
higher. We will discuss in the following
the reason for such discrepancy.

\begin{figure}
\epsfig{file=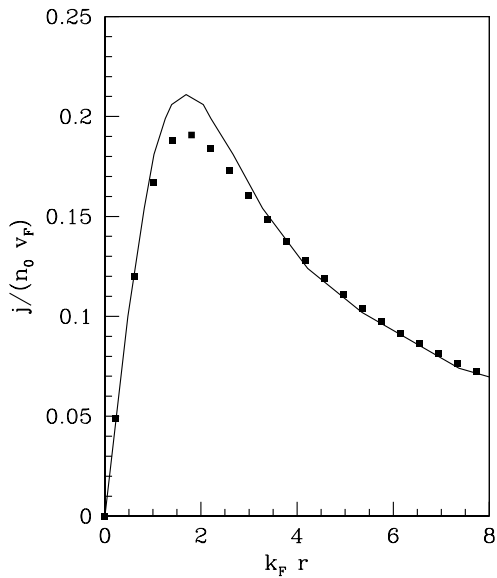,height=3.3 in,clip=}  
\caption{Current density distribution near the vortex core.
Squares: our results.
Solid line: BdG calculations from Ref. \cite{sensarma}.}
\label{fig6}
\end{figure}

We computed the superfluid current density ${\bf j}=\rho {\bf v}$ 
circulating around the vortex core (see Fig. (\ref {fig6})), and found
a peak value at $r_{max}$, in a remarkable overall agreement with
the T=0 Bogoliubov-De Gennes calculations of Ref. \cite{sensarma}
(shown in Fig. (\ref {fig6}) with a solid line).

We find the agreement with the results of 
Ref. \cite{sensarma} particularly rewarding
since it shows that an important superfluid observable 
related to vorticity can indeed be described 
accurately by our Density Functional approach.

The scale of the maximum circulating current is set by
the critical velocity which is determined 
by pair-breaking on the BCS side
and by collective excitations on the BEC side \cite{sensarma}.
This is why our approach, which cannot
obviously describe single-particle processes
like pair-breaking, is nonetheless able to 
reproduce the current pattern.
For $r<r_{max}$ the kinetic energy cost associated 
with the current flow becomes larger than the
condensation energy. Thus $r_{max}$ gives 
an estimate of the distance from the core center 
below which superfluidity is partially suppressed.

We note that the peak position $r_{max}$ in 
Fig. (\ref {fig6}) coincides with the 
value of $r$ where the vortex velocity field 
becomes equal to the {\it local} sound velocity,
i.e. when 
\beq 
{\hbar \over 2mr}\sim \sqrt{2\alpha n(r) ^{2/3}/3m}
\label{cross}
\eeq 
where $n(r)$ is the vortex core density profile
shown in Fig. (\ref {fig5}). 

In Ref. \cite{Sala_Anci_Jos} we have found that, at unitarity, our approach 
leads to a maximum Josephson current across a barrier which is practically 
the same as the one obtained from BdG. This suggests that, at unitarity, 
the current is limited by Landau's criterion for the creation of collective 
excitations, and not by single particle excitations. It is then not 
surprising that also in the present case our maximum current is close 
to that obtained by a much more demanding microscopic calculation based 
on BdG equations (\cite{sensarma,bulg-yu})

It must be said that in Ref. \cite{sensarma,bulg-yu} the vortex structure 
is imposed on the condensate order parameter $\Delta$. Therefore near the 
core the current goes to zero because $\Delta$ vanishes and the superfluid 
density vanishes with it, since $\rho_s \propto \Delta^2 $. However in the 
core region, where the superfluid velocity exceeds the critical value for 
the creation of single particle excitations,  $\rho_s < \rho$, so that the 
total density is non zero even at the core where $\rho_s=0$.

On the contrary, in our approach which does not take into account single 
particle excitations, all the fluid is superfluid, and therefore 
the vanishing at the core of the superfluid current due to vorticity implies 
that the total density is zero there. 
 
Results similar to those reported above were obtained from the simulations
of the lower density system, the main difference
being the shallower vortex cores (which scale as
$k_F^{-1}$). The calculated superfluid 
current density around the core
of an isolated vortex, plotted as a function of $k_Fr$, is 
indistinguishable from the one shown in Fig. (\ref {fig6}).

\begin{figure}
\epsfig{file=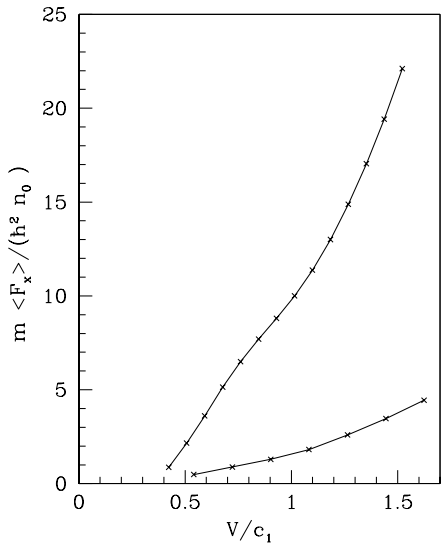,height=3.1 in,clip=}  
\caption{Average drag force for the high and low density systems, 
plotted as a function of $V$.}
\label{fig8}
\end{figure}

We show in Fig. (\ref {fig8}) 
the calculated drag force on the object, as a function 
of the velocity $V$.
Each value is obtained as a time-average \cite{winie}
of curves like the one shown in Fig. (\ref {fig7})
(the average is taken over a time interval where
the drag force has already reached a plateau).
From these results, a value of the critical velocity $v_c\sim 0.4\,c_1$ 
is obtained, in agreement with the simple estimate (\ref{vcrit}).

We also find that the drag force first increases linearly with the velocity
("Stoke's law", usually associated with laminar drag) 
and then turns to a quadratic behavior ("Newton's law", usually
associated with turbulent flow),
since at supersonic velocities there is also a
contribution to the drag associated 
with sound radiation. A similar behavior is observed in the case
of an impenetrable cylinder moving inside a
dilute BEC \cite{winie}.

The phenomenology of the
dissipative motion of an object displaced through a
UFG, as it appears from our calculations, 
is qualitatively similar in many aspects, in spite of the
different non-linear interactions, to the 
behavior observed in the case of an object 
moving in a BEC \cite{winie,frisch}: the occurrence
of vortex emissions in pairs and the associated density patterns
are similar in the two systems, and also the behavior of the 
drag as a function of the object velocity.
There are important differences though:
the emitted vortices in the UFG are doubly quantized, 
as expected from fermion pairing; the predicted 
critical velocity is different; the vortex core structure is
also different, scaling in the present case as the
inverse Fermi momentum.

\section{\bf CONCLUSIONS} 

In conclusion, we have numerically studied the
motion of an object in the ultracold unitary Fermi gas.
We described the system by using an extended density 
functional approach, which has been used recently to
successfully describe a number of static and dynamical 
properties of cold Fermi gases. 

We find that quantized vortices are spontaneously 
generated in pairs during the time evolution at
supercritical velocities. Moreover, the profile of the current density 
as a function of the distance from the core is quantitatively 
close to the one found by the much more demanding solution of the BdG 
solutions. We explain this agreement by observing that at distances 
larger than the one corresponding to the maximum current density, 
in both treatments the 
superfluid density coincides with the total density, since the speed 
of the fluid is below its critical value for the creation of single 
particle excitations. 
At shorter distances, in the core region, the current density is again 
similar in the two treatments: in both it vanishes implying that the 
superfluid density is zero at the center of the vortex. The main 
difference is that while in our case the superfluid density 
coincides with the total density, in the BdG treatment the density 
may have a contribution from a "normal" component related to single 
particle excitations and this component provides a nonzero density 
at the center. 

We acknowledge Fondazione CARIPARO (project of excellence 
"Macroscopic Quantum Properties of Ultracold Atoms under 
Optical Confinement") and University of Padova 
(progetto di ateneo "Quantum Information with Ultracold Atoms 
in Optical Lattices") for partial support. 

\end{document}